\documentclass[preprint,aps,12pt,preprintnumbers,eqsecnum,nofootinbib]{revtex4}
\usepackage{graphicx}
\usepackage{subfigure}
\usepackage{epstopdf} 
\usepackage{braket}
\usepackage{amssymb,amsmath,amsfonts,comment,latexsym,graphicx,epstopdf,bbold,float}
\usepackage{color}
\usepackage{fancyvrb}
\usepackage{chngcntr}
\usepackage{etoolbox}
\usepackage{bbm,epsfig}
\usepackage{array}
\usepackage{hepunits}

\newcommand{\be}{\begin{equation}}
\newcommand{\ee}{\end{equation}}
\newcommand{\ba}{\begin{eqnarray}}
\newcommand{\ea}{\end{eqnarray}}

\newcommand{\grts}{\raise.3ex\hbox{$>$\kern-.75em\lower1ex\hbox{$\sim$}}}
\newcommand{\lets}{\raise.3ex\hbox{$<$\kern-.75em\lower1ex\hbox{$\sim$}}}

{\catcode`\|=\active\gdef\Braket#1{\left<\mathcode`\|"8000\let|\bravert {#1}\right>}}

\def\bravert{\egroup\,\vrule\,\bgroup}

\usepackage{subfigure}

\usepackage{color}
\usepackage{amssymb,amsmath,amsfonts}
\usepackage{epstopdf} 
\usepackage{braket}

\begin{document}
%
%  
% Title of paper
\title{\vspace*{0.5in} 
Composite Gravitons \\ from Metric-Independent Quantum Field Theories\footnote{Invited Brief Review for Modern Physics Letters A.}
\vskip 0.1in}
\author{Christopher D. Carone}\email[]{cdcaro@wm.edu}

\affiliation{High Energy Theory Group, Department of Physics,
William \& Mary, Williamsburg, VA 23187-8795}

\date{\today}
\begin{abstract}
We review some recent work by Carone, Erlich and Vaman on composite gravitons in metric-independent quantum field theories, 
with the aim of clarifying a number of basic issues.  Focusing on a theory of scalar fields presented previously in the literature, we 
clarify the meaning of the tunings required to obtain a massless graviton.  We argue that this formulation can be interpreted as the 
massless limit of a theory of massive composite gravitons in which the graviton mass term is not of Pauli-Fierz form.  We then suggest 
closely related theories that can be defined without such a limiting procedure (and hence without worry about possible ghosts).  
Finally, we comment on the importance of finding a compelling ultraviolet completion for models of this type, and discuss some 
possibilities.
\end{abstract}

\maketitle

\section{Introduction} \label{sec:intro}
A quantum field may represent a fundamental degree of freedom, or a composite state described within the context of an effective 
theory that is applicable in a low-energy regime. The idea that the gauge fields of the standard model may not be fundamental has appeared 
periodically in the literature over the past 50 years~\cite{Bjorken:1963vg,Terazawa:1976xx,Suzuki:2016aqj}.   Such models must have a gauge 
invariance even when no gauge fields are included {\em ab initio}.  Of particular interest to us here are models like those described in 
Ref.~\cite{Suzuki:2016aqj} where the gauge invariance is maintained by adopting rather odd-looking actions that are non-polynomial in the 
fields.  It is then shown that the massless gauge field emerges as a composite state, demonstrated by studying the properties of appropriately 
chosen scattering amplitudes.  A theory whose Lagrangian has no gauge field but nonetheless produces one in the infrared appeals to a 
certain sense of theoretical frugality; this motivates further study of these and other similar models, even if their starting point 
is rather unconventional. Moreover, models in which all the force carriers are emergent in this way may relate the values of their couplings 
in the ultraviolet to the cut off of the theory, potentially leading to interesting relations~\cite{Terazawa:1976xx}.   This may serve as a separate motivation.

The present mini-review describes quantum field theory models of this type, proposed by Carone, Erlich and Vaman~\cite{Carone:2016tup,Carone:2017mdw}, that are designed to produce an emergent graviton in the infrared\footnote{For other quantum field theory approaches to gravity, see Ref.~\cite{new}, and references therein.}.  We will focus primarily on a model of scalar fields in this review, since all the relevant features of this  framework can be illustrated with the fewest technical complications.  A similar theory involving fermionic constituents (that is somewhat more complicated to analyze) was presented in Ref.~\cite{Carone:2018ynf}.   While models of composite gauge fields, like the one of Ref.~\cite{Suzuki:2016aqj}, have gauge-invariant actions that do not include a fundamental gauge field, the theories we consider here, by analogy, have actions that are generally covariant but that do not include a fundamental graviton field.  Hence, they are metric independent.  Early work on theories where the graviton arises as a composite state can be found in 
Refs.~\cite{Akama:1978pg,Akama:1977hr,Amati:1981rf,Amati:1981tu,Wetterich:2003wr}. The starting point for our discussion will be the 
metric-independent action
\begin{equation}
S=\int d^Dx\ \left(\frac{\tfrac D2-1}{V(\phi^a)} \right)^{\frac{D}{2}-1}
\sqrt {\bigg|\det \left(\sum_{a=1}^N \partial_\mu\phi^a \,\partial_\nu\phi^a 
+\sum_{I,J=0}^{D-1}\partial_\mu X^I \,\partial_\nu X^J\, \eta_{IJ}\right)\bigg|} \,\,\, .
\label{eq:scalars}
\end{equation}
This is a theory of $N+D$ real scalar fields, with the $X^I$ fields distinguished since they will be gauge fixed so that
\begin{equation}
X^I=  x^\mu\delta_\mu^I \ \sqrt{\frac{V_0}{\tfrac{D}{2}-1}-c_1} , \ \ I=0,\dots,D-1 \, . \label{eq:staticgauge}
\end{equation}
Here $V_0$ has been chosen to be the same as the constant part of the potential 
\begin{equation}
V(\phi^a) = V_0 + \Delta V-c_2 \,\,\, . \label{eq:thepot}
\end{equation} 
Note that we may define the constant part of the potential in terms of the two constants $V_0$ and $c_2$, and the overall scale of the
$X^I$ profile in terms of the two constants $V_0$ and $c_1$, without any loss of generality.  The counterterms $c_1$ and $c_2$ are chosen to 
normal order the operators $\partial_\mu \phi^a \partial_\nu\phi^a$ and $m^2 \phi^2$, respectively ({\em i.e.}, the counterterms will cancel the loop formed by contracting the two $\phi^a$ fields in either operator).   This choice will allow the resummation of scattering diagrams to all orders 
in $1/V_0$  but at leading order in $1/N$, which will allow us to confirm a massless spin-2 pole, for a special choice of $V_0$.    We review this analysis in Sec.~\ref{sec:pole}.   In Sec.~\ref{sec:interp}, we clarify the meaning of the tuning of $V_0$, as well as its appearance in both Eqs.~(\ref{eq:staticgauge}) and (\ref{eq:thepot}).  We will argue that the formulation presented in Refs.~\cite{Carone:2016tup,Carone:2017mdw} is interpreted consistently as the massless limit of a massive composite gravity model, where the graviton mass does not have the Pauli-Fierz form.  Since this implies the presence of ghosts, we discuss how that can be avoided by an alternative formulation in Sec.~\ref{sec:nolimit}.   We discuss possible ultraviolet (UV) completions for this kind of model in Sec.~\ref{sec:uv}, and summarize our conclusions in Sec.~\ref{sec:conc}.

\section{Graviton Pole} \label{sec:pole}
\begin{figure}[b]
\centerline{\includegraphics[width=4.0in]{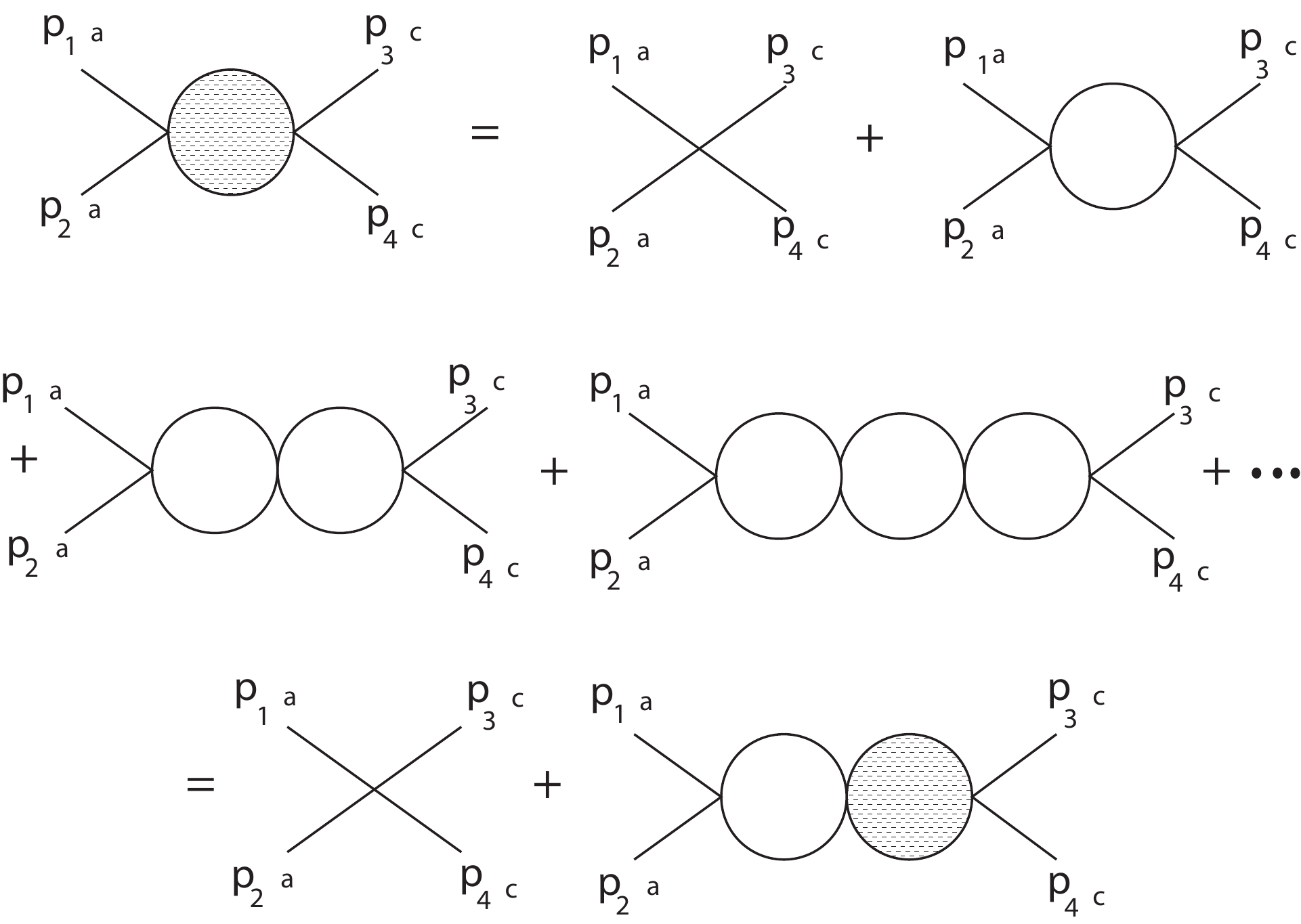}}
\vspace*{8pt}
\caption{Two-into-two scalar scattering, $a \,\, a \rightarrow c\,\, c$, with $a\neq c$.\protect\label{fig:one}}
\end{figure}

In this section, we review the approach of Refs.~\cite{Carone:2016tup,Carone:2017mdw} towards analyzing the theory defined in
Eq.~(\ref{eq:scalars}).  We first assume, for simplicity, the potential of an O($N$)-symmetric free scalar field theory
\begin{equation}
\Delta V(\phi^a)=\sum_{a=1}^N\frac{m^2}{2}\phi^a\phi^a.
\end{equation}
We then expand Eq.~(\ref{eq:scalars}) formally in powers of $1/V_0$.   The counterterms $c_1$ and $c_2$, which would appear multiplying a variety of local operators, can be omitted from that expansion, since any interaction in which they appear will (by design) exactly cancel another in which a scalar loop is closed by contracting a pair of  $\phi^a$ on a factor of $\partial_\mu \phi^a \partial_\nu\phi^a$ or $m^2 \phi^2$.  
One then finds
\begin{equation}
S=\int d^Dx \left[ \frac{V_0}{D/2-1}+\frac{1}{2}\sum_{a=1}^N \partial_\mu \phi^a \partial^\mu \phi^a -\Delta V(\phi^a ) + {\cal L}_{\rm int} \right]
\,\,\, ,
\label{eq:expanded}
\end{equation}
where, after some algebra, ${\cal L}_{\rm int}$ takes the remarkably simple form
\begin{equation}
{\cal L}_{int} =-\frac{1}{4 V_0} {\cal T}_{\mu\nu}\,  \Pi^{\mu\nu|\alpha\beta}\, {\cal T}_{\alpha\beta} + {\cal O}(1/V_0^2) \,\,\, .
\label{eq:lint}
\end{equation}
Here, ${\cal T}^{\mu\nu}$ is the energy-momentum tensor for the free scalar fields $\phi^a$ in Minkowski space
\begin{equation}
{\cal T}^{\mu\nu}= \sum_{a=1}^N\bigg[\partial^\mu \phi^a \partial^\nu \phi^a - \eta^{\mu\nu} \left(\frac{1}{2} \partial^\alpha \phi^a \partial_\alpha \phi^a - \frac{1}{2} m^2 \phi^a \phi^a \right)\bigg]  \,\,\, ,
\label{eq:tflat}
\end{equation}
and $\Pi^{\mu\nu|\alpha\beta}$ represents the tensor structure
\begin{equation}
\Pi^{\mu\nu|\alpha\beta} = \frac{1}{2}  \left[ \left(\frac{D}{2}-1\right) \, ( \eta^{\mu\alpha}\eta^{\nu\beta}+\eta^{\mu\beta}\eta^{\nu\alpha} ) - \eta^{\mu\nu} \eta^{\alpha\beta} \right]   \,\,\,.
\end{equation}
Unlike Eq.~(\ref{eq:scalars}), this action is suitable for conventional diagrammatic study.    In particular, it was argued in Refs.~\cite{Carone:2016tup,Carone:2017mdw} that the two-into-two scattering of scalars of type $a$ into type $c$, with $a \neq c$, and at leading order in $1/N$, gives the set of scattering diagrams shown in Fig.~\ref{fig:one}, which can be exactly resummed.
As indicated by the diagrams in the bottom line of Fig.~\ref{fig:one}, the scattering amplitude $ i {\cal M}$ 
has a convenient recursive representation.  We define    
\begin{equation}
i {\cal M} (p_1,a \,; p_2,a \rightarrow p_3,c\,; p_4,c)\equiv E_{\mu\nu}(p_1,p_2) [i \,A^{\mu\nu|\rho\sigma}(q) ] \,E_{\rho\sigma}(p_3,p_4)\,\,\, ,
\end{equation}
where the factors of $E_{\mu\nu}$ represent the Feynman rule for the flat-space energy-momentum tensor in Eq.~(\ref{eq:tflat}),
\begin{equation}
E_{\mu\nu}(p_1,p_2) \equiv - (p_1^\mu \, p_2^\nu + p_1^\nu \, p_2^\mu) + \eta^{\mu\nu} (p_1 \cdot p_2 +m^2) \,\,\, .
\label{eq:exln}
\end{equation}
The amplitude of Fig.~\ref{fig:one} may then be written   
\begin{equation}
A^{\mu\nu | \rho\sigma} = A_0^{\mu\nu|\rho\sigma} + {K^{\mu\nu}}_{\alpha\beta} \, A^{\alpha\beta | \rho\sigma} \,\,\, ,
\label{eq:recurse}
\end{equation}
where $A_0$ represents the tree-level diagram
\begin{equation}
A^{\mu\nu|\rho\sigma}_0 = - \frac{1}{4 V_0} \left[ (\tfrac D2-1) \, \left(\eta^{\nu\rho} \eta^{\mu\sigma}+ \eta^{\mu\rho} \eta^{\nu\sigma}\right) - \eta^{\mu\nu} \eta^{\rho\sigma} \right]  \,\,\, ,
\label{eq:tamp}
\end{equation}
and the kernel ${K^{\mu\nu}}_{\alpha\beta}$ is given by
\begin{equation}
{K^{\mu\nu}}_{\alpha\beta} = -\frac{1}{2} \lambda \left[1-\frac{D}{12} \frac{q^2}{m^2} \right] (\delta^\mu_\alpha \delta^\nu_\beta+
\delta^\mu_\beta \delta^\nu_\alpha)  + {\cal O}(q^4) \,\,\, ,
\label{eq:kernel}
\end{equation}
where $q=p_1+p_2=p_3+p_4$ and 
\begin{equation}
\lambda \equiv \frac{N(D/2-1)}{2 V_0} \frac{\Gamma(-D/2)}{(4\pi)^{D/2}} (m^2)^{D/2} \,\,\, .
\end{equation}
Here we use dimensional regularization, with $D=4-\epsilon$, to evaluate the loops, since this choice preserves general covariance.  
We comment further on the choice and meaning of the regulator in Sec.~\ref{sec:uv}. For the choice of $V_0$ for which $\lambda=-1$, the 
momentum independent terms in Eq.~(\ref{eq:recurse}) cancel and one can solve for the amplitude\footnote{Our definition of the 
counterterms $c_1$ and $c_2$ leads to the set of diagrams shown in Fig.~\ref{fig:one}.  Other definitions would lead to
additional diagrams that may make resummation of the scattering amplitude intractable.  If such a resummation could be completed, one would anticipate that the final results would remain unchanged, given a tuning equivalent  to the one in our present scheme.} in the vicinity of $q^2=0$,
\begin{equation}
A^{\mu\nu|\rho\sigma}(q) = -\frac{M_{{\rm Pl}}^{2-D}}{D-2} \, \left[(\tfrac D2-1) \,( \eta^{\nu\rho} \eta^{\mu\sigma} + \eta^{\nu\sigma} \eta^{\mu\rho}) 
- \eta^{\mu\nu} \eta^{\rho\sigma} \right] \, \frac{1}{q^2} \,\,\, ,
\label{eq:ampres}
\end{equation}
where $M_{\rm{Pl}}$ is the $D$-dimensional Planck mass, 
\begin{equation}
M_{{\rm Pl}}= m\,\left[\frac{N \, \Gamma(1-\frac{D}{2})}{6\, (4 \pi)^{D/2} }\right]^{1/(D-2)} \,\,\, . \label{eq:mpscat}
\end{equation}
Eq.~(\ref{eq:mpscat}) implies that we must take $\epsilon$ to be small and negative, which further implies that $V_0$ is positive.  The 
amplitude in Eq.~(\ref{eq:ampres}) displays the correct tensor structure for the gauge-invariant part of the propagator of a massless, 
spin-two field.   A similar diagrammatic study of two-into-four scattering in the same theory was employed to study the self-interactions of the 
composite graviton state and the results were found to be consistent with Einstein gravity, up to higher-derivative corrections~\cite{Carone:2017mdw}.   We do not summarize that more burdensome calculation here, but refer the reader to the original literature.  It is worth noting that the scalar mass $m$ in Eq.~(\ref{eq:mpscat}) is smaller than $M_{{\rm Pl}}$ by a factor that is proportional to $\sqrt{N/|\epsilon|}$, which we assume is large.  For example, a TeV-scale scalar mass would result from $N/|\epsilon| \sim {\cal O}(10^{35})$.  In this sense, our toy model  serves as a possible analogy to more realistic scenarios:  the $\phi^a$ fields play the role of ordinary matter (which is light) and generate their own graviton state (which is massless), with couplings that are Planck-suppressed.  One might also imagine realistic models in which the $\phi^a$ are included in addition to other matter fields in the theory, with phenomenological consequences.  We comment  on the approach one might take to construct more realistic models in Sec.~\ref{sec:uv}.  

Finally, it is worth stressing that Eq.~(\ref{eq:mpscat}) requires that $\epsilon$ is finite.  Unlike a conventional renormalizable 
field theory, where one would take $\epsilon$ to zero, and absorb divergences in a finite number of couplings (which is not the case here), the finite regulator implies that loop diagrams in the present framework give finite and calculable radiative corrections that depend on the assumed form of the tree-level theory.  Similar statements could be made had we chosen Pauli-Villars fields as regulators instead:  the value of the Planck and Pauli-Villars scale are related, so that the Pauli-Villars states may not be decoupled from the theory.  They would also lead to finite radiative corrections in any loop 
calculation of interest.  The use of dimensional regularization with finite $\epsilon$, rather than Pauli-Villars fields at a fixed mass scale, is a choice motivated by convenience; either would act as a proxy for the generally covariant physics that completes the theory in the UV.   What that physics may be is an issue we return to in Sec.~\ref{sec:uv}.

\section{Interpretation and Subtleties} \label{sec:interp}

In this section, we discuss a number of subtleties related to the formulation of the model that were not fully addressed in the original literature.

\subsection{The meaning of the tunings}
\begin{figure}[h]
\centerline{\includegraphics[width=3.0in]{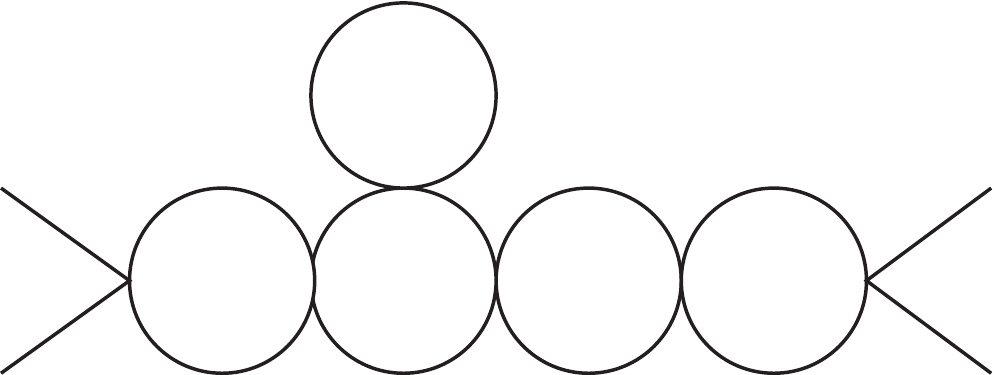}}
\vspace*{8pt}
\caption{Example of a diagram that is exactly cancelled by the counterterms $c_1$ and $c_2$.\protect\label{fig:two}}
\end{figure}

We reviewed in the previous section the method of establishing the existence of a massless graviton.   The result in Eq.~(\ref{eq:ampres}) required that we
set $\lambda=-1$, or equivalently, tune
\begin{equation}
V_0 = - \frac{N(D/2-1)}{2} \frac{\Gamma(-D/2)}{(4\pi)^{D/2}} (m^2)^{D/2}  \,\,\, .
\label{eq:v0val}
\end{equation}
To understand the physical meaning of this choice, let us first compute the counterterms $c_1$ and $c_2$, which were designed to eliminate an infinite number of 
scalar loop tadpole diagrams like the one shown in Fig.~\ref{fig:two}, which are also of leading order in $1/N$.   This choice led us to a tractable calculation.   A straightforward calculation, using the same regulator, yields at leading order in $1/N$
\begin{equation}
c_1 = - \frac{N}{2}  \frac{\Gamma(-D/2)}{(4\pi)^{D/2}} (m^2)^{D/2} \,\,\,\,\, \mbox{ and } \,\,\,\,\,
c_2 = \frac{N}{2}  \frac{\Gamma(1-D/2)}{(4\pi)^{D/2}} (m^2)^{D/2} \,\,\, .
\label{eq:cvals}
\end{equation}
One notices that $V_0 / (D/2-1) - c_1$ exactly vanishes with this tuning, a fact that was missed in Ref.~\cite{Carone:2016tup} due to a sign error, 
but corrected in Ref.~\cite{Carone:2017mdw}.  This quantity is precisely what determines the assumed vacuum expectation value (vev) for the fields $X^I$ in Eq.~(\ref{eq:staticgauge}).  A set of $D$ scalar fields with the profile given in Eq.~(\ref{eq:staticgauge}) have appeared elsewhere in the literature, namely in models which implement a gravitational Higgs mechanism, for the purpose of producing massive gravitons~\cite{tHooft:2007rwo}.   From that perspective, the meaning of the tuning of $V_0$ is demystified: it is the one value that avoids the spontaneous breaking of general covariance so that we obtain a massless spin-2 pole in the solution of Eq.~(\ref{eq:recurse}).

This observation, however, leads to some interesting complications.   The profile of the $X^I$ field in Eq.~(\ref{eq:staticgauge}) is achieved via a gauge-fixing, analogous to the static gauge fixing condition in string theory, but not if the profile is exactly vanishing.  A natural way around this problem is to implement the gauge fixing by working in the limit
\begin{equation}
\frac{V_0}{D/2-1} - c_1 \longrightarrow 0
\end{equation}
without actually hitting the limit point.  This will assure that the massless graviton identified in Sec.~\ref{sec:pole} still remains in the spectrum of the theory. 

On the other hand, this approach may assure that other unwanted states are present as well.  If one repeats the calculation of Sec.~\ref{sec:pole} away from 
the limit point, one finds
\begin{equation}
A^{\mu\nu|\rho\sigma}(q) = \frac{1}{\lambda}\frac{M_{{\rm Pl}}^{2-D}}{D-2} \, \left[(\tfrac D2-1) \,( \eta^{\nu\rho} \eta^{\mu\sigma} + \eta^{\nu\sigma} \eta^{\mu\rho}) 
- \eta^{\mu\nu} \eta^{\rho\sigma} \right] \, \frac{1}{q^2 - \frac{12}{D \lambda} m^2 (1+\lambda)} \,\,\, ,
\label{eq:ampres2}
\end{equation}
which reproduces Eq.~(\ref{eq:ampres}) in the limit $\lambda \rightarrow -1$.   Provided $(1+\lambda)/\lambda >0$, there is a 
(non-tachyonic) massive graviton pole.  While a massive spin-$2$ state has additional degrees of freedom, the tensor structure of 
the amplitude remains unchanged, and there is no van-Damn-Veltman-Zhakarov (vDVZ) discontinuity~\cite{vanDam:1970vg,Zakharov:1970cc}.   
This result can be understood if the graviton mass is not of Pauli-Fierz form, which implies the existence of a spin-$0$ ghost~\cite{VanNieuwenhuizen:1973fi}.  
In particular,  given a mass term of the form
\begin{equation}
{\cal L}_{\mbox{mass}} = -\frac{m_g^2}{2} (h^{\mu\nu}\, h_{\mu\nu} - \frac{1}{2} \, h^2) \,\,\, ,
\end{equation}
the spin-$2$ and ghost degrees of freedom are degenerate in mass~\cite{Hinterbichler:2011tt} and their propagators can be combined, so that the 
parts which contribute to our scattering amplitude have precisely the tensor structure shown in Eq.~(\ref{eq:ampres2}).  This can 
be confirmed using the formula in Ref.~\cite{Park:2010zw}  for the graviton propagator assuming an arbitrary non-Pauli-Fierz mass term.  So, we may 
still have a massless composite graviton, but at the expense of a composite ghost. 

Does this ghost cause any problems in the limit that the graviton mass is taken to zero?   The situation is unclear.  Certainly the diagrams that 
we have considered in Sec.~\ref{sec:pole} are unaffected since the composite ghost only appears after summing internal chains of scalar loops 
which also yield the massive composite graviton; one of the degrees of freedom of the massive graviton exactly cancels the ghost (as we described 
earlier), while two degrees of freedom do not couple to the conserved energy-momentum tensors on the external lines.  What is left are the two 
degrees of freedom we usually associate with a massless graviton.  We can then smoothy take the massless limit of the amplitude considered in 
Sec.~\ref{sec:pole}, recovering our earlier results.  However, the fact that the spectrum contains a stable ghost state suggests that the exact S-matrix
of the theory is not unitary and raises the worry that some violation of unitarity at the perturbative level lurks somewhere at higher loop order in the 
theory.\footnote{Here we assume that a violation of unitarity above $M_{\mbox{Pl}}$ due to our choice of regulator would be addressed in a UV-complete
theory, while the problem originating from ghosts would likely persist in such a completion.}  Rather than attempt to resolve these issues, we will suggest 
a modification of the model that avoids these potential problems entirely. We describe this in Sec.~\ref{sec:nolimit}.   

\begin{figure}[h]
\centerline{\includegraphics[width=4.0in]{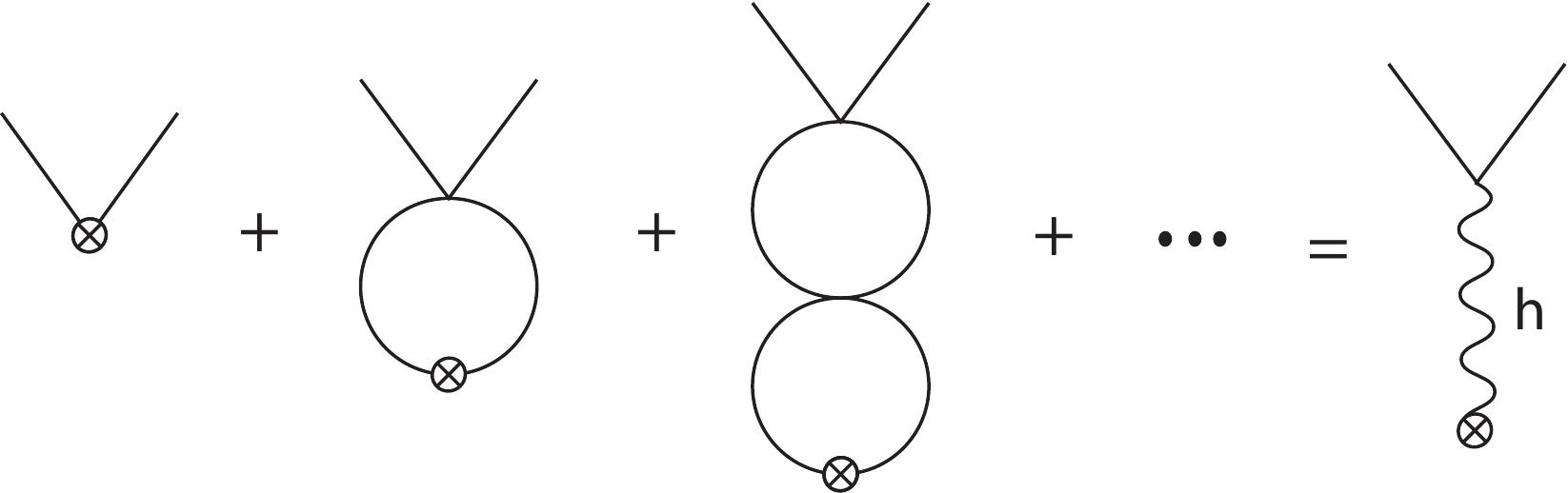}}
\vspace*{8pt}
\caption{Composite graviton tadpole for $\Delta c_2 \neq 0$.\protect\label{fig:three}}
\end{figure}

However, before we discuss that reformulation, let us discuss the remaining tuning that was incorporated in the formulation of Sec.~\ref{sec:pole}; 
the appearance of the same constant $V_0$ in the potential and in the $X^I$ vev.   To understand the meaning of this choice, let us imagine shifting 
the value of $V_0$ in the potential, which we can equivalently interpret as a shift $c_2 \rightarrow c_2 - \Delta c_2$, with $c_2$ defined precisely as 
before, a counterterm that effectively normal orders the operator $m^2 \phi^2/2$.   The shift $\Delta c_2$ appears everywhere in the interaction 
Lagrangian where $\Delta V$ appears; hence one finds a new interaction
\begin{equation}
{\cal L}_{int} =-\frac{1}{4 V_0} {\cal T}_{\mu\nu}\,  \Pi^{\mu\nu|\alpha\beta}\, {\cal T}_{\alpha\beta}
- \frac{\Delta c_2}{2 V_0} \eta^{\alpha\beta} {\cal T}_{\alpha\beta} \,\,\, .
\label{eq:lintnew}
\end{equation}
Imagine connecting the interaction proportional to $\Delta c_2$ to a chain of loops, as in Fig.~\ref{fig:three}.  This is precisely the same
diagrammatic resummation as in our scattering calculation except that one external two-scalar-line factor $E(p_1,p_2)_{\mu\nu}$ is replaced 
by $-\Delta c_2 \, \eta_{\mu\nu}$.  This corresponds to scattering off a graviton tadpole, an indication of an instability that one would expect when
doing a flat-space expansion $g_{\mu\nu} = \eta_{\mu\nu} + h_{\mu\nu}$ in the presence of  a cosmological constant.  Hence, $\Delta c_2$ must be 
tuned to zero, which was built into our initial choice for the form of Eqs.~(\ref{eq:staticgauge}) and (\ref{eq:thepot}).  Aside from the vanishing of the
graviton mass, further evidence that our tunings eliminate a cosmological constant was provided in Ref.~\cite{Carone:2017mdw}, where two-into-four scattering 
amplitudes were studies to gain information on the form of the three-composite graviton vertex.  Various contributions to the induced 
vertex that were independent of graviton momenta were found to exactly cancel, just as one would expect if the cosmological constant has been set to zero.
In summary, the two tunings we have identified eliminate the spontaneous breaking of general covariance (by enforcing a vanishing $X^I$ field vev)  and also eliminate the terms one would expect if a cosmological constant were present.  

\subsection{Relation to the Weinberg-Witten Theorem}

The Weinberg-Witten Theorem provides an impediment to constructing a theory of massless composite gravitons if the theory has an 
S-matrix and a non-vanishing, conserved Lorentz-covariant energy-momentum tensor~\cite{Weinberg:1980kq}.   With the usual definition
\begin{equation}
T_{\mu\nu} = \frac{2}{\sqrt{|g|}} \frac{\delta S}{\delta g^{\mu\nu}} \,\,\, ,
\end{equation}
it is clear that  Eq.~(\ref{eq:scalars}) has no conserved energy-momentum tensor, by virtue of its metric independence.  Alternatively, one
can write the action in terms of an auxiliary metric $g_{\mu\nu}$, 
\begin{equation}
S=\int d^Dx\,\sqrt{|g|}\left[\frac{1}{2}g^{\mu\nu}\left(\sum_{a=1}^N\partial_\mu\phi^a\,\partial_\nu\phi^a+\sum_{I,J=0}^{D-1}\partial_\mu X^I\partial_\nu X^J\eta_{IJ}\right)
-V(\phi^a)\right], \label{eq:auxg}
\end{equation}
while imposing a constraint on the partition function that the energy-momentum tensor vanishes.\footnote{Imposing a constraint on the partition function differs from simply integrating over the 
auxiliary metric, since the latter leads to the presence of a functional determinant in the remaining integrand; hence, we cannot show at a quantum 
mechanical level that our formulation is equivalent to simply changing the order of integration over the metric and the fields in a model of induced gravity.}   
 Solving the constraint for $g_{\mu\nu}$ and replacing the auxiliary metric in the action leads precisely to our 
starting point in Eq.~(\ref{eq:scalars})~\cite{Carone:2016tup,Carone:2017mdw}.  One might expect that similar results should be obtained in 
metric-independent theories that cannot necessarily be re-expressed in terms of an auxiliary metric, so the existence of a mapping between these two descriptions 
is not essential.  In any case, the absence of a Lorentz-covariant energy-momentum tensor violates the assumptions of the Weinberg-Witten theorem, so the 
results we have described present no inconsistency.  Note that the vanishing of the energy-momentum tensor of the full theory (gravity and matter) does not 
preclude the graviton from coupling to the non-vanishing energy-momentum tensor of the scalar fields alone, as we have seen in Sec.~\ref{sec:pole}.

Conversely, if one were handed an alternative theory that matched only the leading order terms appearing in the expanded action, 
Eqs.~(\ref{eq:expanded}) and (\ref{eq:lint}), one would expect the Weinberg-Witten theorem to apply, even though our calculation at leading 
order in $1/N$ would give the same results in both theories.  In this case, one simply expects that a graviton mass would appear at next order 
in $1/N$ in the alternative theory; this is precisely the order at which that theory would fail to reproduce the metric-independent and generally 
covariant form of Eq.~(\ref{eq:scalars}).

Based on our results, and the general covariance of Eq.~(\ref{eq:scalars}), we would expect the massless graviton pole to persist away from the 
large-$N$ limit, even if there were a single scalar field $\phi$.  In this case, however, other non-perturbative techniques would be required to 
demonstrate the existence of the massless spin-$2$ state.  This direction is worthy of further investigation.

\section{A Model without the Limit} \label{sec:nolimit}

In Sec.~\ref{sec:interp}, we clarified the interpretation of the results presented in Sec.~\ref{sec:pole}.  The $X^I$ fields are gauge-fixed, with no 
remaining physical degrees of freedom, with a profile that must be taken to approach zero as a limit so that general covariance is not spontaneously 
broken.   One then wonders: are the $X^I$ fields really serving any meaningful purpose in the theory?   In this section, we show that the results of 
Sec.~\ref{sec:pole} could be obtained in the same theory  {\em without} the $X^I$ fields, working at the $\lambda=-1$ limit point where general 
covariance is exact.   In this case, one does not need to worry about the potential consequence of ghosts from the composite massive gravity theory 
that may remain after the massless limit is taken.

The reformulation of the theory can be summarized as follows:  We begin with the simpler action
\begin{equation}
S=\int d^Dx\ \left(\frac{\tfrac D2-1}{V(\phi^a)} \right)^{\frac{D}{2}-1}
\sqrt {\bigg|\det \left(\sum_{a=1}^N \partial_\mu\phi^a \,\partial_\nu\phi^a 
\right)\bigg|} \,\,\, ,
\label{eq:simpler}
\end{equation}
and add zero to it in the following way:
\begin{equation}
S=\int d^Dx\ \left(\frac{\tfrac D2-1}{V(\phi^a)} \right)^{\frac{D}{2}-1}
\sqrt {\bigg|\det \left( \frac{V_0}{\frac{D}{2}-1} \, \eta_{\mu\nu} - c_1 \, \eta_{\mu\nu} + \sum_{a=1}^N \partial_\mu\phi^a \,\partial_\nu\phi^a 
\right)\bigg|} \,\,\, ,
\label{eq:simpler}
\end{equation}
where $V_0$ and $c_1$ have precisely the same values given in Eqs~(\ref{eq:v0val}) and (\ref{eq:cvals}), respectively.  In other words, we now have no $X^I$ fields, we have not specified a gauge fixing condition, and we have simply added {\em zero} to the quantity within the determinant
in Eq.~(\ref{eq:simpler}).  In doing so, the general covariance of the theory remains unaffected.  Nevertheless, we may still organize our perturbative 
expansion as before, formally treating $c_1$ as a counterterm that normal orders $\partial_\mu \phi^a \partial_\nu \phi^a$ and retaining all diagrams that are leading order in a $1/N$ expansion.   This is identical to the calculation presented in Sec.~\ref{sec:pole} and we would again identify a massless graviton pole.   In this case, however, we had no need to spontaneously break general covariance and then take a limit to restore it, allowing us to avoid the potential problems associated with ghosts, as discussed in Sec.~\ref{sec:interp}.

\begin{figure}[t]
\centerline{\includegraphics[width=4.0in]{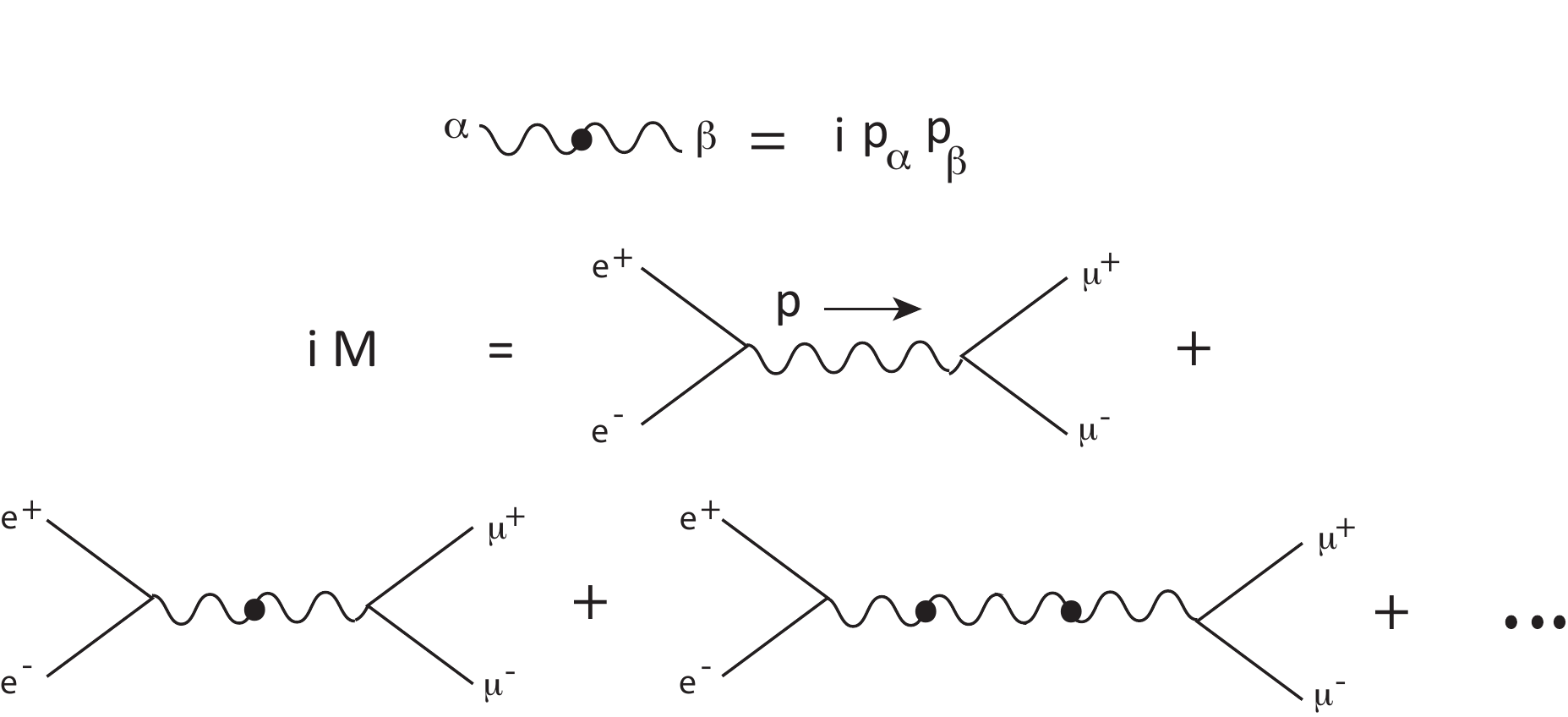}}
\vspace*{8pt}
\caption{QED example discussed in the text.\protect\label{fig:four}}
\end{figure}

One point may be puzzling about this result:  no mention was made of gauge fixing.  While gauge fixing is of course necessary, we need not specify it to establish the existence of the massless pole in the way that the calculation of Sec.~\ref{sec:pole} is organized.   To understand this, it is useful to consider an analogy.  Let us imagine computing $e^+ e^- \rightarrow \mu^+ \mu^-$ scattering in QED in the following way.  Let us add precisely zero to the QED Lagrangian by two cancelling gauge-non-invariant terms
\begin{equation}
{\cal L}_{\mbox{QED}} = \left[{\cal L}_{\mbox{QED}}-\frac{1}{2} (\partial^\mu A_\mu)^2\right] + \frac{1}{2} (\partial^\mu A_\mu)^2 \,\,\, ,
\label{eq:qedfun}
\end{equation}
where ${\cal L}_{\mbox{QED}}$ is the usual QED Lagrangian without any gauge fixing terms. To echo our previous language, we have not specified a gauge fixing condition, but have simple added zero to the original Lagrangian, so that the U(1) gauge
invariance of the theory remains unaffected.   However, we will now organize our perturbation theory so that the last term in Eq.~(\ref{eq:qedfun}) is treated as an interaction (which will have to be formally resummed to all orders so that we obtain a reliable result).  The part of the quantity in brackets that is quadratic in the fields is invertible and leads to the same photon propagator that one would have obtained had we chosen to work in Feynman gauge,
\begin{equation}
\widetilde{D}^{\mu\nu} = \frac{-i g^{\mu\nu}}{k^2+i \epsilon}  \,\,\,.
\end{equation}
With the Feynman rule for the new interaction shown in Fig.~\ref{fig:four}, the scattering amplitude of interest is represented by the sum of diagrams shown in the same figure.  Notice that each diagram involving insertions of the new vertex provides a factor of the s-channel momentum $p^\mu$ that contracts with the conserved fermion current on the external lines.   Hence, each of the diagrams involving a new vertex vanish and we are left with the first diagram, precisely the correct answer we would have obtained in QED had we fixed the gauge at the start.
This result is a consequence both of our unconventional expansion and the particular set of diagrams we have chosen to consider.  The fact that the quadratic terms
in QED are not invertible until a gauge is fixed would be manifest had we summed only the internal lines of the diagrams first, which would have led to a divergent
result.  However, order-by-order in our expansion, the problematic momentum-dependent terms cancel when contracted with the conserved currents on the external 
lines, so that we are left with the correct gauge-invariant part of the amplitude.  Diagrams where this cancelation does not occur involve loops with internal photon lines, but 
these are not relevant if we are only interested in tree-level, s-channel photon exchange.  The analogous situation is obtained in our gravity theory, where the diagrams we 
consider encode tree-level, s-channel composite graviton exchange; momentum dependent terms are discarded when contracted with the conserved energy-momentum 
tensors on the external lines before the infinite series of diagrams is summed.  What is left is the correct gauge-invariant part of the amplitude.  For our present purposes, 
we need not specify the gauge fixing condition, which would certainly be relevant if we were computing amplitudes involving loops of scalar loops ({\em i.e.}, graviton loops), 
where we have no expectation that terms originating from the momentum-dependent parts of the composite graviton propagator would vanish.  How exactly such a gauge fixing condition is implemented will not effect our results, as one can see from the following argument:  in an effective field theory with the composite graviton field $h_{\mu\nu}$, one can directly write down a general gauge-fixed form for the $h_{\mu\nu}$ propagator and observe that the parts which depend on the gauge choice vanish in the particular s-channel amplitude that we study.  This implies that our corresponding result for the amplitude near the graviton pole will not change regardless of how the 
gauge choice is imposed on the fundamental scalar degrees of freedom in the full theory.

\section{UV Completions} \label{sec:uv}

We have commented in Sec.~\ref{sec:intro} that the model reviewed here presents a gravitational analogy to a similar class of composite gauge 
boson models, and is of theoretical interest for similar reasons.  Could a model of this type be developed into a realistic description of nature?   First, a metric-independent field theory that reduces to the standard model plus gravity would need to be presented; this is likely a matter of fleshing out technical details, since the basic conceptual approach is clear from the scalar and fermionic models that have appeared in the literature, as well as the formulation without $X^I$ fields suggested in Sec.~\ref{sec:nolimit} of the present review.  The recipe one would follow is similar to the one suggested in the case of our scalar theory in Ref.~\cite{Carone:2016tup}:  One could construct a generally covariant action, involving the field content of interest, with the help of a non-dynamical auxiliary metric that one immediately eliminates via the constraint of vanishing energy-momentum tensor.  The technical difficulty comes in solving for the auxiliary metric (or vielbein) field in the general case; while this could be done exactly in our scalar theory, and in the particular theory of fermions presented in 
Ref.~\cite{Carone:2018ynf}, more general theories are typically not as cooperative, with solutions only available in the form of a perturbative expansion in the parameter $1/V_0$.   Applying such a perturbative approach to eliminating the auxiliary field leads to an expanded action analogous to Eqs.~(\ref{eq:expanded}) and (\ref{eq:lint}), that would allow diagrammatic studies to confirm the existence of an emergent graviton state.  Generalization to the standard model seems possible via this approach and is the subject of ongoing work.

However, a compelling ultraviolet completion appears to be more challenging.  The use of dimensional regularization in the evaluation of scalar loops in Sec.~\ref{sec:pole} was convenient since it does not break the general covariance of the theory.   The Planck scale was then found to be a function of the scalar mass $m$ and the finite regulator $1/\epsilon$.   Unfortunately, this choice for regulating the theory does not allow us to extrapolate scattering amplitudes above the Planck scale, which we would want in a UV complete theory. (Higher-derivative operators suppressed by the $\epsilon$-dependent Planck scale will lead to scattering amplitudes that grow with energy, leading eventually to a violation of unitarity.) In the model we have discussed, dimensional regularization represents a placeholder for whatever 
generally covariant, physical regulator cuts off the theory.   In a similar way, one could have maintained general covariance with Pauli-Villars regulators, but one would again suffer from the same problem, namely that the unitarity of scattering amplitudes would be violated above a finite Pauli-Villars scale. (In this case, unitarity violation is associated with the fact that Pauli-Villars states have negative norm.)  Another possibility is to develop a lattice formulation of the theory, with a finite lattice spacing setting the Planck scale.  One would not have to worry about scattering amplitudes above the Planck scale, since field theory would not be applicable above the Planck scale.  The difficulty in this approach would be in finding a lattice formulation that preserves a massless graviton without requiring that one go to the continuum limit.   Finally, an even more exotic completion has been suggested by Erlich, in which the critical points of the $X^I$ field configurations are given 
a fundamental role in the stochastic behavior of hidden variables that leads to the emergence of something that approximates quantum theory~\cite{Erlich:2018qfc}.   As discussed earlier, the formulation of the theory with the $X^I$ fields implies the presence of ghosts. It would be desirable to show (or arrange)
that these states lead to no inconsistencies in any UV completion that gives the $X^I$ fields a special significance.

\section{Conclusions} \label{sec:conc}

In this mini-review, we have summarized a recent approach to models of composite gravitons, with a focus on clarifying issues that were not transparent in the original literature.   In particular, we have explained that the original formulation of these models involving so-called 
``clock-and-ruler" fields $X^I$ produces a massive composite graviton, whose mass is taken to approach zero as the expectation values of 
the $X^I$ field are chosen to do the same.  A separate tuning eliminates a linear graviton term which indicates an instability associated with the presence of a cosmological constant.  We clarified the relevance of the Weinberg-Witten theorem, explaining (for pedagogical clarity) why an inequivalent theory consisting only of the interaction terms included in our amplitude analysis (but not our full theory in its entirety), would yield the same results we obtain in scalar-scalar scattering at leading order in a $1/N$ expansion, but would not be expected to preserve a massless graviton pole at higher order.    We also present a reformulation of the model without the $X^I$ fields, not discussed previously in the literature, which preserves general covariance at all times and provides the massless graviton state without the worry about ghosts that might linger 
after taking a limit in a massive composite gravity model.   Finally, we point out that the generalization to the entire standard model plus 
gravity seems to be a matter of resolving technical rather than conceptual difficulties, but that a compelling ultraviolet completion is not yet 
at hand and worthy of further investigation.   The approach summarized here is unconventional.   Nevertheless, the value of thinking about 
an old problem in unconventional ways is that it might suggest new avenues for progress that might not be readily apparent when 
following well-worn paths.   It is hoped that the worked summarized here will, at the very least, have that positive effect.
%\begin{figure}[t]
 %\begin{center}
 %   \includegraphics[width=.5\textwidth]{figDDetectRR.eps}
%   \caption{Allowed range in $m_{S_0}$, consistent with direct detection bounds with the relic
 %  density held fixed at $\Omega h^2 = 0.1186$.  The approximate lower bound on $m_{S_0}$ shown 
 %  on the left corresponds to $m_{\tilde{h}_0}^{min} / 2$, where $m_{\tilde{h}_0}^{min} = 327$~ GeV
  % is the indirect bound from Ref.~\cite{Carone:2009nu} that is discussed in the text.}
%       \label{fig:two}
 % \end{center}
%\end{figure}

%%%%%%%%%%%%%%%%%%%%%%%%%%%%%%%%%%%%%%%%%%%%%%%%%%%%%%
\begin{acknowledgments}  
We thank Josh Erlich and Diana Vaman for many enlightening discussions.  We thank the NSF for support under Grant No. PHY-1819575.  
\end{acknowledgments}
%%%%%%%%%%%%%%%%%%%%%%%%%%%%%%%%%%%%%%%%%%%%%%%%%%%%%%%%%%%     

%\appendix
%\section{} \label{sec:appendix}
    
\end{document}